# The Lieb-Oxford Lower Bounds on the Coulomb Energy, Their Importance to Electron Density Functional Theory, and a Conjectured Tight Bound on Exchange

John P. Perdew[1] and Jianwei Sun[2]

[1]Departments of Physics and Chemistry, Temple University, Philadelphia, PA 19122

[2]Department of Physics and Engineering Physics, Tulane University, New Orleans, LA 70118

*Abstract*: Lieb and Oxford (1981) derived rigorous lower bounds, in the form of local functionals of the electron density, on the indirect part of the Coulomb repulsion energy. The greatest lower bound for a given electron number $N$ depends monotonically upon $N$, and the $N \to \infty$ limit is a bound for all $N$. These bounds have been shown to apply to the exact density functionals for the exchange- and exchange-correlation energies that must be approximated for an accurate and computationally efficient description of atoms, molecules, and solids. A tight bound on the exact exchange energy has been derived therefrom for two-electron ground states, and is conjectured to apply to all spin-unpolarized electronic ground states. Some of these and other exact constraints have been used to construct two generations of non-empirical density functionals beyond the local density approximation: the Perdew-Burke-Ernzerhof (PBE) generalized gradient approximation (GGA), and the strongly constrained and appropriately normed (SCAN) meta-GGA.

1. Applied Mathematics as a Foundation for Theoretical Physics

We would like to begin by congratulating Elliott Lieb, on the occasion of his 90[th] birthday, for a long and important career in applied mathematics, and for his 2022 American Physical Society Medal for Exceptional Achievement in Research, "for major contributions to theoretical physics through obtaining exact solutions to important physical problems, which have impacted condensed matter physics, quantum information, statistical mechanics, and atomic physics".

Applied mathematics and theoretical physics are distinct but intertwined activities. The applied mathematician proves theorems that are demonstrably true under the stated assumptions, but chooses those assumptions and theorems that may be relevant to theoretical physics or other fields. The theoretical physicist explains and predicts what can be found in the physical world, or develops concepts and methods that can be used for that purpose. Both rely on intuition and logic, but the theoretical physicist can also rely on the theorems proved by the applied mathematician.

The field of theoretical physics that we know best is the density functional theory of Hohenberg, Kohn, and Sham [1,2]. Important mathematical contributions to this field have been made by Lieb [3,4], Levy [5], and others including those listed in [6]. In Ref. [4], Lieb studied the mathematical properties of the associated functionals and proposed alternative formulations. Here we want to concentrate on the lower bounds on the Coulomb energy derived by Lieb and Oxford [3], and to explain how they have guided the development of two generations of practical approximations to the density functional for the exchange-correlation energy of a many-electron system, and thus to our quantitative understanding of normal matter.

2. Synopsis of Density Functional Theory

Atoms, molecules, and solids are composites of electrons and nuclei. In atomic units, the electrons have electric charge -1, spin ½ (making them fermions) with z-components $\sigma = +1/2$ or ↑ and -1/2 or ↓, and light unit identical masses which require a quantum mechanical treatment. The nuclei have integer positive charges and much heavier masses that make them almost classical. Under normal conditions, these systems are often close to their ground states or states of lowest energy. The allowed energies (for the nuclei at rest) are eigenvalues of the Hamiltonian

$$\widehat{H} = \sum_{i=1}^{N} -\frac{1}{2}\nabla_i^2 + \sum_{i=1}^{N} v(r_i) + \widehat{V_{ee}} + \frac{1}{2}\sum_\alpha \sum_{\beta \neq \alpha} Z_\alpha Z_\beta / |R_\beta - R_\alpha|. \tag{1}$$

Electron $i$ has three-dimensional position operator $r_i$ and nucleus $\alpha$ has classical position $R_\alpha$. The first term $\widehat{T}$ is a sum of one-electron kinetic energy operators. The second term is a sum of one-electron potential energy operators, usually but not necessarily the Coulomb attraction of the electrons to the nuclei:

$$v(r) = -\sum_\alpha Z_\alpha/|r - R_\alpha|. \tag{2}$$

The third term, which plays a key role in this article, is the potential energy operator for Coulomb repulsion between pairs of electrons:

$$\widehat{V_{ee}} = \frac{1}{2}\sum_{i=1}^{N}\sum_{j\neq i} 1/|r_j - r_i|. \tag{3}$$

The last term, which is just a number, is the Coulomb repulsion between pairs of nuclei.

The mean number of electrons in volume element $d^3r$ is $n(r)d^3r$. Hohenberg and Kohn [1] showed that the ground-state electron density $n(r)$ determines the external potential $v(r)$, and that there exists a universal density functional $F[n]$, such that minimization of

$$E_v[n] = F[n] + \int d^3r\, v(r)n(r) \tag{4}$$

at fixed electron number $N = \int d^3r\, n(r)$ and external potential $v(r)$ yields the ground-state electron density and energy (excluding the last term of Eq. (1)). Kohn and Sham [2] made this approach practical by writing

$$F[n] = T_s[n] + U[n] + E_{xc}[n], \tag{5}$$

where $T_s[n]$ is the ground-state kinetic energy of non-interacting electrons of density $n(r)$,

$$U[n] = \frac{1}{2}\int d^3r\, d^3r'\, n(r)n(r')/|r' - r| \tag{6}$$

is the Hartree electrostatic interaction of the density with itself, and the residue $E_{xc}[n] = E_x[n] + E_c[n]$ is called the exchange-correlation energy. It is often a relatively small contribution to the total energy, but a large contribution to the binding energy of one atom to another, making it "Nature's glue". Then one can in principle find the exact ground-state energy, related energy differences, and the electron density by replacing the cumbersome $N-$ electron Schrödinger equation by $N$ tractable self-consistent one-electron Schrödinger equations:

$$\left[-\frac{1}{2}\nabla^2 + v(r) + \frac{\delta U}{\delta n(r)} + \frac{\delta E_{xc}}{\delta n(r)}\right]\varphi_{a,\sigma}(r) = \epsilon_{a,\sigma}\varphi_{a,\sigma}(r), \tag{7}$$

$$n(r) = \sum_{a,\sigma}|\varphi_{a,\sigma}(r)|^2\, \theta(\mu - \epsilon_{a,\sigma}). \tag{8}$$

$\mu$ is a Lagrange multipler determined by constraining $n(r)$ to integrate to $N$ electrons. The Heaviside step function $\theta$ in Eq. (8) restricts the sum to one-electron states that are occupied in the ground state. The Kohn-Sham one-electron wavefunctions or orbitals $\varphi_{\alpha,\sigma}(r)$ are by Eq. (7) functionals of the electron density $n(r)$. Kohn and Sham [2] also proposed the local density approximation

$$E_{xc}^{LDA}[n] = \int d^3 r\, n(r) \varepsilon_{xc}^{unif}(n(r)), \qquad (9)$$

in which $\varepsilon_{xc}^{unif}(n)$ is the known exchange-correlation energy per electron in an electron gas of uniform density $n$. Levy [5] extended the domain of densities on which the exact functionals are constructed, and gave precise definitions to all the exact functionals. Those definitions involve expectation values of operators using an antisymmetric interacting wavefunction $\Psi_{N,\lambda=1}$ that yields the density $n(r)$ and minimizes the expectation value of $\hat{T} + \widehat{V_{ee}}$, and a non-interacting antisymmetric wavefunction $\Psi_{N,\lambda=0}$ that yields the same density and minimizes the expectation value of $\hat{T}$. $\Psi_{N,\lambda=0}$ is typically a single Slater determinant of the occupied Kohn-Sham orbitals $\varphi_{\alpha,\sigma}(r)$. Here $\lambda$ is the coupling constant that scales the physical electron-electron interaction. Levy's "constrained search" over all antisymmetric wavefunctions yielding a given density $n(r)$ makes it easy to generalize from density ($n$) to spin-density ($n_\uparrow, n_\downarrow$) functional theory.

Density functional theory is formally exact for the ground-state energy and density of a system with the Hamiltonian of Eq. (1). There are in fact several exact variants of the theory, depending in part on how the ground-state density is defined, and some of those are more suitable to accurate and computationally efficient approximation than others. For example, using the separate spin densities $n_\uparrow(r)$ and $n_\downarrow(r)$ instead of the total density $n(r)$ provides more information to an approximation and makes it more accurate, even at the level of the local spin density approximation. Constructing the density from wavefunctions or pure states also provides more information than using one from ensembles or mixed states, and it is now clear that advanced approximations are more accurate for the former case [7].

After several generations of refinements beyond the LDA of Eq. (9) and its spin-density generalization, approximate spin-density functionals can now more accurately predict what atoms, molecules, and solids can exist, and with what properties. The most predictive functionals are constructed by satisfying known exact constraints: mathematical properties that have been derived for the exact functional $E_{xc}[n]$. Among these are lower bounds based on the work of Lieb and Oxford [3].

3. Synopsis of the Lieb-Oxford Lower Bounds on the Coulomb Energy

The expectation value of the electron-electron repulsion for an $N$-electron wavefunction of density $n(r)$ can be written as the sum of a positive direct or Hartree term and an indirect term $I[\Psi_N]$:

$$< \Psi_N | \widehat{V_{ee}} | \Psi_N > = U[n] + I[\Psi_N] \tag{10}$$

with $U[n]$ defined in Eq. (6). The wavefunction $\Psi_N$ is not restricted to be a ground-state, and it can be anti-symmetric or symmetric; in fact, the expectation value can be taken in an ensemble of wavefunctions. There is no upper bound on $I[\Psi_n]$, but there is a negative greatest lower bound [3] that depends on $N$ (but not on the spin of the electron):

$$I[\Psi_N] \geq -C_N \int d^3 r \, n^{\frac{4}{3}}(r). \tag{11}$$

The optimal constant $C_N$ increases with $N$:

$$C_1 \leq C_2 \leq \cdots \leq C_\infty, \tag{12}$$

where $C_1 = 1.092$, $C_2 \geq 1.234$ (or 1.256 [8]), and $C_\infty \leq 1.68$. Chan and Handy [9] improved the last bound slightly to $C_\infty \leq 1.64$. The bounds are much tighter and thus more useful than the earlier [10] $C_\infty \leq 8.52$. There is thus a greatest lower bound independent of $N$:

$$I[\Psi_N] \geq -C_\infty \int d^3 r \, n^{\frac{4}{3}}(r). \tag{13}$$

The significance for density functional theory is that the Lieb-Oxford bounds are all local functionals of the density, like the LDA itself and relevant to more advanced approximations. But the indirect part of the Coulomb interaction enters density functional theory only indirectly. From the adiabatic connection fluctuation dissipation theory [11,12], we find [13,14]

$$0 \geq E_x[n] = I[\Psi_{N,\lambda=0}] \geq E_{xc}[n] = \int_0^1 d\lambda \, I[\Psi_{N,\lambda}] \geq I[\Psi_{N,\lambda=1}] \geq -C_N \int d^3 r \, n^{4/3}(r), \tag{14}$$

where $\Psi_{N,\lambda}$ is now that *N*-electron antisymmetric wavefunction that yields the density $n(r)$ and minimizes the expectation value of $\hat{T} + \lambda \widehat{V_{ee}}$, providing a continuous connection between the non-interacting and interacting systems. The exchange-correlation energy of density functional theory is an integral over the coupling constant $\lambda$ from 0 to 1, which includes a positive kinetic energy of correlation in addition to the negative indirect Coulomb energy at the physical λ= 1. Thus Eq, (14) yields lower bounds on the exchange energy functional and on the exchange-correlation energy functional.

For comparison,

$$0 \geq E_x^{LDA}[n] = -0.739 \int d^3r\, n^{\frac{4}{3}} \geq E_{xc}^{LDA}[n] > \sim -1.43 \int d^3r\, n^{\frac{4}{3}} \geq -C_\infty \int d^3r\, n^{4/3}, \qquad (15)$$

obeys the Lieb-Oxford bound for a system with an arbitrarily large number of electrons, because LDA is by construction exact for an electron gas of uniform density. The lower bound on $E_{xc}^{LDA}[n]$ in Eq. (15) arises from the low-density limit of the uniform gas correlation energy per electron, as parametrized (accurately but not exactly) in Ref. [15] using a formula from Ref. [16] and quantum Monte Carlo data from Ref. [17]. In fact, LDA inherits several exact constraints from its exactness for an infinite class of uniform densities, explaining its better-than-expected performance for real systems.

The Lieb-Oxford bounds for $N \gg 1$ are not expected to be close unless the electron-electron Coulomb correlation is strong. In the low-density limit of the uniform electron gas, the electrons are perfectly correlated, forming a body-centered cubic Wigner crystal that minimizes the expectation value of the Coulomb repulsion energy. Thus Perdew [13] and Levy and Perdew [18] conjectured that this limit provides the optimal $C_\infty \approx 1.43$, where the numerical value comes from the fit discussed in the previous paragraph; a more precise 1.44 comes from the energy of the Wigner crystal, but the difference is negligible for the construction of approximate functionals. Lewin and Lieb [19] derived a tight bound $C_{UEG} \approx 1.45$ for the uniform electron gas, but suggested that the combination of surface effects with long-range interactions might rule out the equivalence between the energies per electron of the infinite Wigner crystal and of the ground-state of a large finite jellium in the low-density limit. That equivalence was later proved rigorously [20,21,22]. Some mathematical properties of the uniform electron gas have been derived in Ref. [23].

For the generalization of the Wigner crystal to the description of strong correlation in inhomogeneous electron densities, see Refs. [24,25].

4. Tight Bound on the Exchange Energy of a Two-Electron Ground State, and Its Conjectured Generalization

.     The exact exchange energy in density functional theory is

$$E_x = I[\Psi_{N,\lambda=0}] = -\frac{1}{2}\sum_\sigma \int d^3r \int d^3r' |\rho_\sigma(r,r')|^2/|r'-r|, \qquad (16)$$

where the one-particle density matrix of Kohn-Sham orbitals of spin $\sigma$ is

$$\rho_\sigma(r,r') = \sum_a \varphi^*_{a,\sigma}(r')\varphi_{a,\sigma}(r)\theta(\mu - \epsilon_{a,\sigma}). \qquad (17)$$

The diagonal of Eq. (17) is the electron spin density $n_\sigma(r)$. Apart from the small differences between the Kohn-Sham and Hartree-Fock orbitals, the exchange energy defined by Eqs. (16) and (17) is just the Hartree-Fock exchange energy of the system. Because the Kohn-Sham orbitals are functionals of the density, so is the Kohn-Sham exchange energy. This exchange energy has coordinate-scaling equalities that the correlation energy does not have, so each has to be approximated separately. For a system of many electrons, the Lieb-Oxford lower bound does not seem to be tight for the exchange energy, or even for the exchange-correlation energy except possibly for strongly-correlated systems. In the two-electron case, however, the sum in Eq. (17) has only one term, making

$$|\rho_\sigma(r.r')|^2 = n_\sigma(r')n_\sigma(r). \qquad (18)$$

The two-electron ground state is spin-unpolarized $(n_\uparrow = n_\downarrow = \frac{n}{2})$, so its exchange energy is

$$E_x^{N=2}[n] = 2E_x^{N=1}[n/2] \geq 2(-1.092)\int d^3r \left(\frac{n}{2}\right)^{4/3} = -0.867\int d^3r\, n^{4/3}, \qquad (19)$$

where we have used the optimal Lieb-Oxford lower bound for the xc = x energy of a one-electron density (which also follows from the earlier work of Gadre, Bartolotti, and Handy [26]). Eq. (19) is a very tight lower bound for the exchange energies of compact spherical two-electron densities (e.g., the He atom), for which it is almost an equality. For the spin-symmetry-unbroken (unpolarized) hydrogen molecule $H_2$, the lower bound of Eq. (19) is very tight at the equilibrium bond length, but it becomes much more negative than the exact exchange energy as the bond is stretched and the density becomes more lobed [27].

The bound of Eq. (19) was derived by Perdew, Ruzsinszky, Sun, and Burke [28], who conjectured that it might provide a lower bound on the exchange energy of a spin-unpolarized density of *any* electron number $N$. No counter-example is known to the authors, and the strongly-constrained and appropriately normed (SCAN) meta-GGA [29] for the exchange-correlation energy, based in part on that conjecture, has had remarkable successes.

Because the spin-density functional for the exact exchange energy obeys a spin-scaling relation [30]

$$E_x[n_\uparrow, n_\downarrow] = \frac{1}{2}\{E_x[2n_\uparrow] + E_x[2n_\downarrow]\}, \tag{20}$$

we only need to approximate $E_x[n]$ for spin-unpolarized densities. Unlike the Lieb-Oxford bounds, the greatest lower bound on the exchange energy depends upon the spin quantum number (1/2 for electrons) and on the relative spin polarization $\frac{n_\uparrow - n_\downarrow}{n}$.

From Eq. (16), an obvious upper bound on the electronic exchange energy is zero, but that limit is reached only when the density tends to zero everywhere. Appendix A argues that there is no tight upper bound of Lieb-Oxford form. A *rigorous* tight lower bound like Eq. (19) for spin-unpolarized densities at *all* electron numbers would be of great value for the construction of constraint-based density-functional approximations, because it would constrain what is typically the largest part of the approximated exchange-correlation energy.

5. Importance of the Lower Bounds on the Indirect Coulomb Energy as Exact Constraints for Density Functional Approximations

Kohn and Sham [2] constructed the local density approximation of Eq. (9) to be exact for slowly-varying inhomogeneous densities; a proof of exactness is given in Ref. [31]. The next step might be expected to be the second-order gradient expansion [32,33],

$$E_{xc}^{GE2}[n] = \int d^3r \, n\varepsilon_{xc}^{unif}(n) + \int d^3r \, \{C_x + C_c(n)\}|\nabla n|^2/n^{4/3}. \tag{21}$$

This expression is asymptotically correct in the limit of densities that vary slowly over three-dimensional space, but it actually worsens the predictions of LDA for real systems, because this truncated expansion does not inherit many of the exact constraints satisfied by LDA. The known gradient coefficient for exchange, $C_x$, is negative, so that lower bounds on the exchange energy are violated by Eq. (21) for densities that vary sufficiently rapidly. That is not such a serious problem for real systems, but it is still one that needs to be corrected. More seriously, the known gradient coefficient for correlation, $C_c(n)$, is positive and large, leading to incorrectly positive correlation energies for real systems. The exact constraints satisfied by LDA can be restored, and others can be satisfied, by generalized gradient approximations (GGAs):

$$E_{xc}^{GGA}[n] = \int d^3r \, F_{xc}^{GGA}(n,s) n \varepsilon_x^{unif}(n), \tag{22}$$

where

$$\varepsilon_x^{unif}(n) = -0.739 n^{1/3} \tag{23}$$

is the exact exchange energy of Eq. (16) per electron in a non-interacting electron gas of uniform spin-unpolarized density (already used in Eq. (15)) and

$$s = 0.1616|\nabla n|/n^{4/3}. \tag{24}$$

The GGA enhancement factor over local exchange can be written as [13,14,34]

$$F_{xc}^{GGA}(n,s) = F_x^{GGA}(s) + F_c^{GGA}(n,s). \tag{25}$$

From Eqs. (14), (15), (20), and (21), the GGA exchange enhancement factors can be constructed to satisfy

$$1.68/[0.739 \times 2^{\frac{1}{3}}] = 1.804 \geq F_x^{GGA}(s) \geq F_x^{GGA}(s=0) = 1. \tag{26}$$

The Perdew-Wang 1991 (PW91) [13,14] GGA and the widely-used Perdew-Burke-Ernzerhof 1996 (PBE) [34] GGA were constructed to satisfy the bounds

$$E_x^{GGA}[n_\uparrow, n_\downarrow] \geq E_{xc}^{GGA}[n_\uparrow, n_\downarrow] \geq -1.68 \int d^3r \, n^{4/3}. \tag{27}$$

as well as other exact constraints. However, the final inequality in Eq. (27), approached by PBE exchange energy at large $s$, is not very important for the exchange energies of most real atoms, molecules, and solids, where the energetically-important regions have $0 \leq s < 3$.

Above the first (LDA) and second (GGA) rungs of the ladder of density functional approximations [35] is the third or meta-GGA (MGGA) rung, which depends upon the Kohn-Sham kinetic energy density $\tau(r)$, e.g., in the SCAN [29] MGGA (based in part on the conjectured tight bound of Eq. (34)),

$$E_{xc}^{MGGA}[n] = \int d^3r \, F_{xc}^{MGGA}(n, s, \alpha) n \varepsilon_x^{unif}(n). \tag{28}$$

where

$$\alpha = \frac{\tau - \tau_W}{\tau_{unif}} \geq 0, \tag{29}$$

$$\tau = \frac{1}{2} \sum_{a\sigma} |\nabla \varphi_{a,\sigma}(r)|^2 \theta(\mu - \epsilon_{a,\sigma}), \tag{30}$$

$$\tau_W = \frac{|\nabla n|^2}{8n}, \tag{31}$$

$$\tau_{unif} = 2.871 n^{\frac{5}{3}}. \tag{32}$$

The exchange enhancement factor becomes $F_x^{MGGA}(s, \alpha)$:

$$0.867/0.739 = 1.174 = F_x^{MGGA}(s = 0, \alpha = 0) \geq F_x^{MGGA}(s, \alpha). \tag{33}$$

$\alpha = 0$ recognizes two-electron spin-unpolarized ground states, or more generally regions of space in which a single orbital shape is dominant, $\alpha \approx 1$ with $s \ll 1$ recognizes slowly-varying densities for which GGA can be accurate, and $\alpha \gg 1$ recognizes regions of space in which density tails overlap. Comparison of Eqs. (26) and (33) shows that SCAN exchange is substantially different from PBE exchange. The SCAN [29] meta-GGA is substantially more accurate [36,37,38] than the PBE GGA, in part because its exchange energy satisfies the conjectured tight bound for spin-unpolaized densities at all electron number of section 4,

$$E_x[n] \geq -0.867 \int d^3r \, n^{4/3}. \tag{34}$$

That accuracy gives us extra confidence that the bound of Eq. (34) is exact or nearly exact for all $\alpha$, and not just for $\alpha = 0$. In fact, the SCAN exchange energy is closest to its lower bound of Eq. (34) when $\alpha = 0$, as shown in Fig. 1 of Ref. [29]. Greater computational efficiency and slightly greater accuracy is achieved by the smoother r²SCAN [39] meta-GGA, which satisfies 16 of SCAN's 17 exact constraints, including Eq. (34).

For the $H_2$ molecule with spin-symmetry breaking, which localizes the ↑ and ↓ electrons on different nuclei at large bond lengths, the SCAN exchange-correlation functional is accurate at all bond lengths [40].

There are of course other reasons for the success of SCAN, including its satisfaction of all 17 known exact constraints that a meta-GGA can satisfy (listed in the supplementary information of Ref. [29]), and its fitting to non-bonded "appropriate norms" such as the uniform electron gas and some atoms [29]. Since density functionals are primarily used to predict how atoms bond together, LDA, PBE, and SCAN, which are not fitted to any bonded system, are regarded as "non-empirical functionals".

Rigorous proof of tight bounds like the conjectured lower bound of Eq. (34) on the exact exchange energy of a spin-unpolarized electron density for all electron numbers would be a valuable contribution to density functional theory.

Note added in proof: Lewin, Lieb, and Seiringer [41] have tightened the bound of Eq. (27), replacing 1.68 by 1.25, which leads to a factor of $1.25/2^{\frac{1}{3}} = 0.99$ in Eq. (34), and is thus close to the conjectured bound of Eq. (34).

*Acknowledgments:* Thanks to Aaron Kaplan, Codina Cotar, Paola Gori-Giorgi, Kieron Burke, and Mathieu Lewin for manuscript suggestions. JPP acknowledges support from the National Science Foundation under Grant No. DMR-1939528, CMMT, with a contribution from CTMC. JS acknowledges the support from the National Science Foundation under Grant No. DMR-2042618.

Appendix A: Upper Bound on the Exact Exchange Energy

From Eq. (16), the exact exchange energy $E_x[n]$ of a spin-unpolarized electronic system has an upper bound of zero, which is achieved only when $n \to 0$. Here we will show that there is no upper bound of the form

$$E_x[n] \leq -C \int d^3 r \, n^{4/3} \qquad (35)$$

for any $C > 0$.

Consider the non-uniform density scaling [34] in Cartesian coordinates

$$n(x, y, z) \to \gamma n(\gamma x, y, z), \tag{36}$$

which leaves the electron number

$$N = \int dx\, dy\, dz\, n(x, y, z) \tag{37}$$

unchanged. In the $\gamma \to \infty$ limit, a three-dimensional density is collapsed to two dimensions, and

$$\int dx\, dy\, dz\, n^{\frac{4}{3}}(x, y, z) \to \gamma^{\frac{1}{3}} \int dx\, dy\, dz\, n^{\frac{4}{3}}(x, y, z), \tag{38}$$

which diverges to ∞. But the exact exchange energy per electron, $E_x[n]/N$, must approach the finite (Eq. (45) of Ref. [42]) and negative definite exchange energy per electron of the two-dimensional system ([43], numerical evidence for a slab model ). That would not happen if Eq. (38) were true for any $C > 0$.

For a slab of uniform electron density and periodic boundary conditions in the $xy$ plane, with the width in the $z$ direction collapsing to zero around the plane $z = 0$, one can use the definition of Eq. (16) and the separability of the Kohn-Sham orbitals to show on one page that the exact exchange energy per unit area approaches that of a truly two-dimensional uniform electron gas with the same number of electrons per unit area. In this limit, the LDA and PBE exchange energies per electron diverge, while SCAN has a qualitatively (but not quantitatively) correct finite limit [44] by virtue of its non-uniform scaling constraint.

The conclusion of this Appendix is consistent with exact and LSDA exchange energies for one-electron densities of increasing nodedness [27].

*References*

[1] P. Hohenberg and W. Kohn, Inhomogeneous electron gas. *Phys. Rev.* **136** (1964), no. 3B, B864-B871.

[2] W. Kohn and L.J. Sham, Self-consistent equations including exchange and correlation effects. *Phys. Rev.* **140** (1965), no. 4A, A1133-A1138.

[3] E.H. Lieb and S. Oxford, Improved lower bound on the indirect Coulomb energy. *Int. J. Quantum Chem.* **19** (1981), 427-439.